\documentclass[twocolumn,showpacs,preprintnumbers]{revtex4}%
\usepackage{epsfig}
\usepackage{graphicx}
\usepackage{dcolumn}
\usepackage{bm}
\usepackage{amsmath}
\usepackage{amsfonts}
\usepackage{amssymb}
\usepackage{color}%
\DeclareMathOperator{\Acos}{acos}

\begin{document}
\title{Non-ergodicity of the motion in three dimensional steep repelling dispersing potentials.}
\author{Anna Rapoport}
\email{anna.rapoport@weizmann.ac.il}
\author{Vered Rom-Kedar}
\email{vered.rom-kedar@weizmann.ac.il}
\affiliation{Weizmann Institute of Science, Israel}
\date{\today}


\begin{abstract}
It is demonstrated numerically that smooth three degrees of
freedom Hamiltonian systems which are arbitrarily close to three
dimensional strictly dispersing billiards (Sinai billiards) have
islands of effective stability, and hence are non-ergodic. The
mechanism for creating the islands are corners of the billiard
domain.

\end{abstract}
\maketitle

%05.20.Dd Statistical mechanics

%45.20.Jj Lagrangian and Hamiltonian mechanics

%05.45.Pq Chaos: numerical simulations

%05.45.-a Nonlinear dynamics and nonlinear dynamical systems

%\preprint{APS/123-QED}

%PACS, the Physics and Astronomy
%Classification Scheme.
\textbf{The motion of a point particle travelling with a constant speed inside
a region $D\in\mathbb{R}^{N}$, }$N\geq2$,\textbf{ undergoing elastic
collisions at the regions's boundary, is known as the billiard problem. Since
the days of Boltzmann, scientists have been using various billiard models to
approximate the classical and semi-classical motion in systems with steep
potentials (e.g. for studying classical molecular dynamics, cold atom's motion
in dark optical traps and microwave dynamics). The invalidity of this
approximation near certain types of trajectories is the main issue of this
paper. Indeed, we examine this approximation in the most robust case of a
scattering Sinai billiard (all the boundary components of the billiard are
smooth, dispersing, and their intersections are all oblique). Such billiards
are known to be ergodic, hyperbolic and strongly mixing, thus small smooth
deformations of the billiard boundaries do not change these properties.
Nonetheless, it had been longed conjectured that by introducing smooth steep
potentials which are close to the billiards, hyperbolicity may be destroyed.
In the two-dimensional settings, it had been proven analytically\textit{ }that
tangent periodic orbits and certain corners produce stability islands for
arbitrarily steep potentials, with precise estimates of the scaling of the
islands size with the steepness parameter. Direct generalization of these
results to higher dimensions may produce non-hyperbolic behavior, but one
would intuitively suspect that in the scattering case there will be always
some unstable directions which will destroy stability. Here, we provide a
mechanism for the creation of islands of effective stability (destroying both
hyperbolicity and ergodicity) in the higher dimensional setting. \ We
demonstrate numerically that the islands of stability are created for
arbitrarily steep potential in \emph{both two and three} dimensional
billiards. Furthermore, we show that the islands are created for an interval
of steepness parameters, hence, for a fixed geometry, one may destroy an
island by either making the potential steeper \emph{or} softer.}

\section{Introduction}

%The behavior of a point particle travelling with a
%constant speed in a region $D\in\mathbb{R}^{N}$, undergoing
%elastic collisions at the regions's boundary, is known as the
%billiard problem. When $D$ is a compact domain and all its
%boundary components are smooth, dispersing, and their
%intersections are all oblique, the resulting dynamics is known to
Sinai billiards are known to be ergodic and strongly mixing
\cite{Sina70,SiCh87,GaOr74}. In many applications
\cite{Gut90,Smil95,kfad01,FGK92} the billiard's flow is a simplified model
which imitates the conservative motion in a steep potential:
\begin{equation}
H=\sum_{i=1}^{N}\frac{p_{i}^{2}}{2}+W(q;\epsilon),\text{ \ \ \ \ }%
W(q;\epsilon)\underset{\epsilon\rightarrow0}{\rightarrow}\left\{
\begin{array}
[c]{cc}%
0 & q\in D\backslash\partial D\\
c & q\in\partial D
\end{array}
\right.  \label{hamiltonian}%
\end{equation}
where $c$ may be infinite. Here we always take the particle's
energy, $h$, to be smaller than $c$ so that the particle is
confined to $D$. An important question is whether the billiard and
the smooth flows are similar for sufficiently small $\epsilon$ --
in particular -- whether the billiard's ergodicity property is
preserved. A definite answer to such a question requires a well
defined limiting procedure \cite{Mar68,RKTu99}. For finite-range
axis-symmetric potentials it was shown that some configurations
remain ergodic \cite{Sina63,Ku76,DoLi91,BaTo04}, while other
configurations may possess stability islands \cite{Bal88,Do96}.
Recently, it was established that in the most general
two-dimensional settings of dispersing billiards (not necessarily
axis-symmetric nor of finite range) the answer is definitely
negative; it was proved that there are two mechanisms for the
creation of stability islands for arbitrarily small $\epsilon$.
One mechanism is a tangency -- periodic orbits or homoclinic
orbits which are tangent to the billiard's boundary produce
islands \cite{RKTu99}. Another mechanism are corners -- a sequence
of regular reflections which begins and ends in a corner (termed a
\emph{corner polygon}) may, under some prescribed conditions,
produce stable periodic orbits \cite{turk03}. In both cases it was
shown that a two-parameter family of potentials
$W(q;\epsilon,\alpha)$ ($\epsilon$ is the softness parameter and
$\alpha$ is responsible for a regular continuous change of the
billiard's geometry) possesses a wedge in the $(\epsilon,\alpha
)$-plane, at which the Hamiltonian flow has an elliptic periodic
orbit. This orbit limits to the tangent billiard orbit/ the corner
polygon as $\epsilon\rightarrow0$. Furthermore, a method for
estimating the width of the stability wedge in the parameter space
and of the area of the elliptic islands in the phase space was
developed; for typical potentials both quantities have a power-law
dependence on $\epsilon$ \cite{RKTu99,turk03}. These findings were
realized experimentally using cold atoms in atom-optics billiards
\cite{kfad01}. In the experiments, a mixing billiard domain is
drawn by a fast moving laser beam which encloses cold atoms. A
small gap is opened after an initial run time, and the fact that
the decay rate of the remaining atoms depends on the gap location
demonstrates that the dynamics is not mixing and that some of the
particles are trapped in stability islands. The numerical
simulations of the experiments show that islands are indeed
produced by corner polygons \cite{kfad01}.

Much less is known on the dynamics in multi-dimensional billiards
( $N\geq3$). Motivated by the Boltzmann hypothesis regarding the
ergodicity of hard sphere gas, the ergodicity property of
hard-wall semi-dispersing billiards were extensively studied (see
\cite{KSSz92,SiSz99,Sim04} and reference there in). Nowhere
dispersing ergodic billiards in $\mathbb{R}^{N}$ with $N\geq3$
were constructed in \cite{Wo90,BuRe97,BuRe98,BuRe98f}. In these
papers and in \cite{ZaSt92} examples of three-dimensional
semi-focusing billiards with mixed phase space were presented.
Conditions under which multi-dimensional billiards with finite
range spherically symmetric potentials are hyperbolic were found
in \cite{BaTo06}. A semiclassical study of  three-dimensional
Sinai billiard was presented in \cite{PrSm00}. Recently, the
asymptotic expansion of regular (non-tangent, away from corners)
motion in steep multi-dimensional potentials by integrals along an
auxiliary multi-dimensional billiard were developed \cite{RRT06}.
In this work the geometry is arbitrary, and error bounds on the
billiard approximation are found.

Here, we demonstrate numerically, for the first time, that islands
of stability are created for arbitrarily small $\epsilon$ in
\emph{both two and three} dimensional soft billiards. The ability
to locate small islands of stability in the six dimensional phase
space of the highly chaotic nearly-billiard $3$ d.o.f. flow may
appear to be hopeless. Three technical innovations enable us to
establish these results numerically. The first idea is to
construct a simple symmetric billiard, so that instead of looking
for islands of stability in arbitrary places, we may concentrate
on the properties of a simple periodic trajectory which exists for
all small $\epsilon$ values by symmetry. We examine its stability
properties by computing the monodromy matrix of the local return
map near this orbit. Inspired by \cite{kfad01,turk03}, we choose a
trajectory which limits, as $\epsilon \rightarrow0$, to the
simplest possible corner polygon - a cord which enters a corner
(see the bold lines in FIG.\ref{fig: 2d geometry} and
FIG.\ref{fig: 3d z geometry}). Furthermore, in the three
dimensional case, by the symmetry of the constructed billiard, the
two non-trivial pairs of eigenvalues of the monodromy matrix are
identical, and are thus controlled by a single parameter. The
second idea is that by using proper rescaling it is possible to
integrate numerically the equations of motion for arbitrarily
small $\epsilon$. Indeed, if we fix the geometry and take small
$\epsilon$ values we encounter the usual problem of stiffness near
the boundary. On the other hand, the equivalent increase of the
billiard domain by a similarity factor does not introduce a
serious numerical problem since $\nabla W$ is small in the
domain's interior. The third idea is that the boundaries of the
wedges of stability in the parameter space may be found
numerically by a continuation scheme on the critical eigenvalues
value. Thus the stability regions may be found effectively and
efficiently.

\section{Billiard geometry}

\begin{figure}[ptb]
\includegraphics[width=7cm]{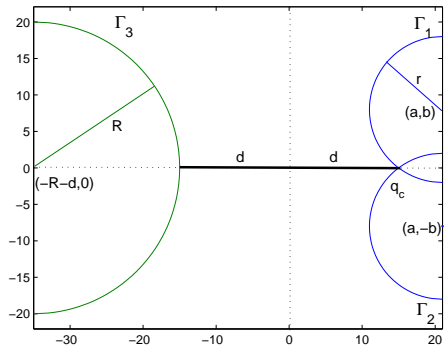}
%Here is how to import EPS art
\caption{(Color online)The billiard geometry in the 2D case. A cord $\gamma$
is denoted by the bold line.}%
\label{fig: 2d geometry}%
\end{figure}

To construct concrete examples, we define the billiard domains as
the region exterior to several spheres $\Gamma_{k}$ with centers
at $A^{k}$ and radii $r^{k}$:
$\Gamma_{k}(A^{k},r^{k})=\{q\in\mathbb{R}^{N}:{\displaystyle\sum
\limits_{i=1}^{N}}(q_{i}-A_{i}^{k})^{2}=\left(  r^{k}\right)
^{2}\},\text{ }N=2\text{ or }3.$ For the two dimensional case we
take three circles (FIG.\ref{fig: 2d geometry}). The first two
circles $(A^{1,2},r^{1,2})=(a,\pm b,r)$ intersect at the point
$q_{c}=(d,0),$ where $d(a,b,r)=a-\sqrt {r^{2}-b^{2}}$ and the
third circle, which has a larger radius, has
$(A^{3},r^{3})=(-R-d(a,b,r),0,R)$ with $R\gg r\geq b$. The angle
between the tangents to the two circles at $q_{c}$ is given by:
\begin{equation}
\alpha_{2D}=\pi-\Acos(1-2\frac{b^{2}}{r^{2}}),\label{alpha}%
\end{equation}
so that when $r=b$ these circles are tangent and $\alpha_{2D}=0$.
The cord $\gamma=\left\{  (x,y)|x\in(-d,d),y=0\right\}  $ is a
\emph{corner polygon}: at $(x,y)=(-d,0)\ $it reflects from the
large circle $\Gamma_{3}$ according to the billiard's reflection
law ($\phi_{in}=\phi_{out}=\pi/2$) and at $(x,y)=(d,0)$ it enters
a corner. We will study the behavior of the smooth system near
this corner polygon, thus the closing of the billiard domain away
from this line is irrelevant here. It may be achieved by a union
of a finite number of dispersing smooth boundaries which meet at
non-zero angles, or by enclosing the whole system in a large box.
For all $\alpha>0$ the family of billiard tables thus defined
belong to the class of Sinai billiards - they are mixing dynamical
systems, having one ergodic component and a positive Lyapunov
exponent for almost all initial conditions.

\begin{figure}[ptb]
\includegraphics[width=7cm]{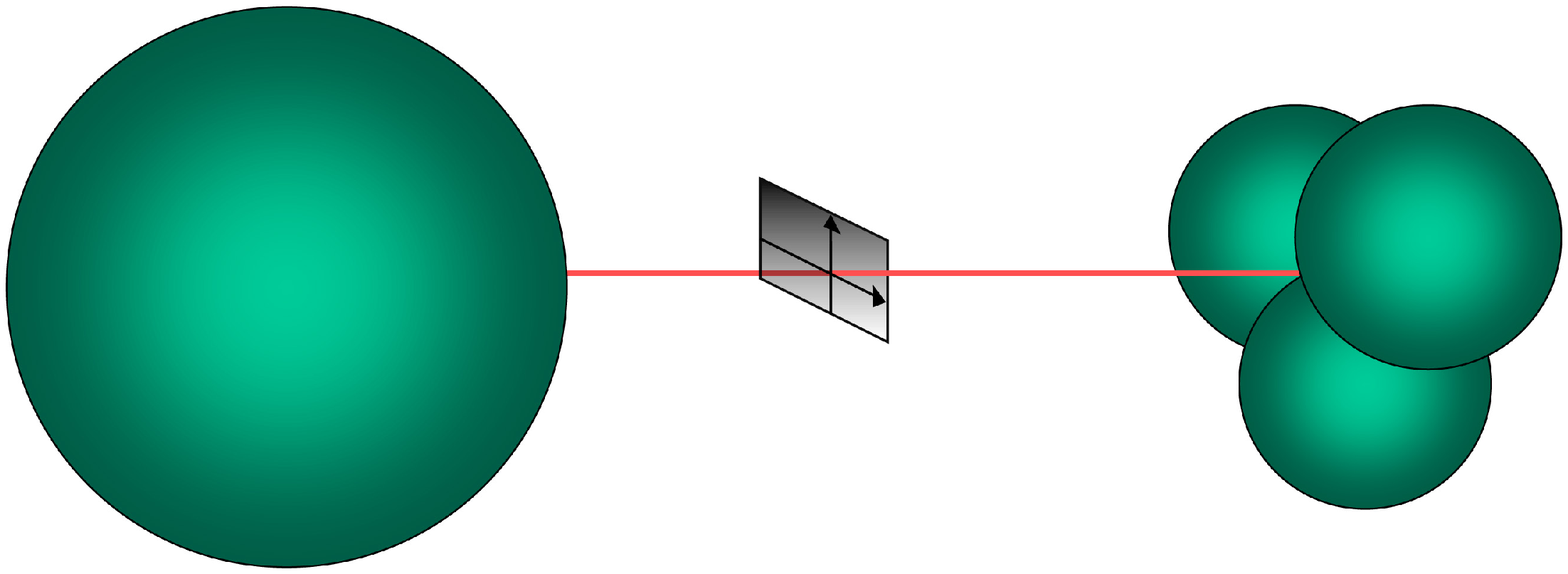}
%Here is how to import EPS art
\caption{(Color online)The billiard geometry in the 3D case. The cord $\gamma$
is denoted by the solid line.}%
\label{fig: 3d_geometry}%
\end{figure}

\begin{figure}[ptb]
\includegraphics[width=7cm]{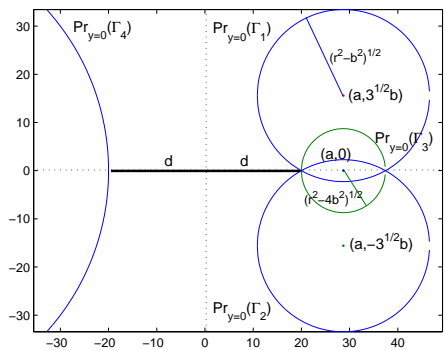}
%Here is how to import EPS art
\caption{(Color online)The billiard geometry in the 3D case, at the cross
section $y=0$. The cord $\gamma$ is denoted by the bold line.}%
\label{fig: 3d z geometry}%
\end{figure}

\begin{figure}[ptb]
\includegraphics[width=7cm]{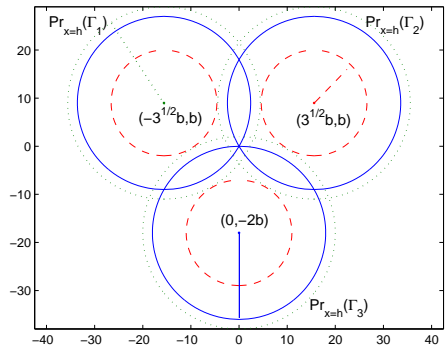}
%Here is how to import EPS art
\caption{(Color online) The billiard geometry in the 3D case: at the
cross-section $x=x_{f}$ the radius of the circles is $r_{f}=\sqrt{r^{2}%
-(x_{f}-a)^{2}}$. Dotted line: $x_{f}=a$, solid line: $x_{f}=d$, dashed line:
$d<x_{f}<a+r$. }%
\label{fig: 3d xa geometry}%
\end{figure}

Similarly, in the three-dimensional case, we take four spheres (FIG.
\ref{fig: 3d_geometry},\ref{fig: 3d z geometry},\ref{fig: 3d xa geometry}).
Three spheres have equal radii $r$ and have equidistant centers:
$(A^{1,2},r^{1,2})=(a,b,\pm\sqrt{3}b,r),$ $(A^{3},r^{3})=(a,-2b,0,r)$. These
three spheres intersect, for $r\geq2b$, at $q_{c}=(d,0,0)$ where
$d(a,b,r)=a-\sqrt{r^{2}-4b^{2}}$. The fourth sphere, of radius $R\gg r$, is
located at a distance $2d$ from the corner point: $(A^{4},r^{4}%
)=(-R-d(a,b,r),0,0,R)$. The angle between the pairs of tangent lines to the
circles of intersections of pairs of spheres is:
\begin{equation}
\alpha_{3D}=\Acos(-\frac{1}{2}(1+\frac{3}{(3-r^{2}/b^{2})}))\label{alpha 3D}%
\end{equation}
so $r=2b$ corresponds to the case $\alpha_{3D}=0$. Furthermore, the cord
$\gamma=\left\{  (x,y,z)|x\in(-d,d),y=z=0\right\}  $ is a corner polygon. Here
again we can close the billiard domain by adding a finite number of dispersing
surfaces which intersect each other in finite angles, or by a large box, so
that for all $\alpha>0$ the resulting billiard domain is compact and
dispersing. Note that if we rescale all the spheres and the distances between
them by a fixed scale $L$, the billiards geometry will not change and the
corresponding corner angles remain unchanged.

\section{ Equations of motion for the smooth flow}

Consider the smooth motion in this region which is induced by the potential
$W(q;w_{0})=\sum_{k=1}^{n}V_{k}(q;w_{0})$; $V_{k}(q;w_{0})$ may be taken as
the Gaussian potential associated with the boundary component $\Gamma_{k}$:
$\ $$V_{k}(q;w_{0})=V(Q_{k}(q);w_{0})=\exp\left(  -\frac{Q_{k}^{2}(q)}%
{w_{0}^{2}}\right)  $, where $Q_{k}(q)$ is the distance between $q$ and the
circle $\Gamma_{k}:\quad$$Q_{k}(q)=\sqrt{{\displaystyle\sum\limits_{i=1}^{N}%
}(q_{i}-A_{i}^{k})^{2}}-r^{k}$ and $w_{0}$ is the softness parameter. In the
cold atom experiment $w_{0}$ corresponds to the width of the laser beam
\cite{kfad01}, and $V(Q_{k}(q);w_{0})$ corresponds to the averaged effective
Gaussian potential which bounds the atoms. Previously, we established that as
this potential tends to a hard wall potential ($w_{0}\rightarrow0$), regular
reflections of the smooth flow tend to those of the billiard
\cite{RKTu99,RRT05}. By the symmetric placement of the spheres, it is clear
that for any $w_{0}<w_{0}^{\ast}$ (where $\min_{\gamma}W(q;w_{0}^{\ast})=h$),
there exists a periodic solution $\gamma(t,w_{0})=(x(t,w_{0}),0,0)$ which
limits, as $w_{0}\rightarrow0$ to the corner polygon $\gamma$. Notice that
studying this system for a fixed $w_{0}$ and a billiard domain which is
increased proportionally by a factor $L$ (so $(A^{k},r^{k})\rightarrow
(LA^{k},Lr^{k})$), is equivalent to studying it in a fixed geometry with
$w_{0}$ replaced by $\epsilon=w_{0}/L$. Thus, by increasing the domain size we
may approach the limit $\epsilon\rightarrow0$ without the numerical problems
associated with the stiff limit $w_{0}\rightarrow0$.

\begin{figure}[ptb]
\includegraphics[height=6.5cm]{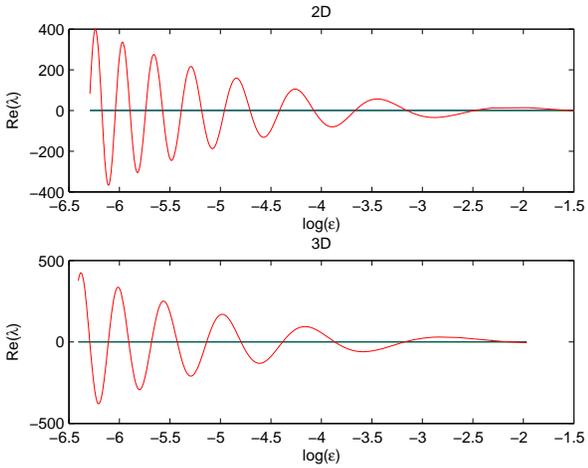}
%\caption[]{2D: Wedges of stability in the parameter space}
\caption{(Color online) The real part of eigenvalue $\lambda$ at $\alpha=0$ as
a function of $log(\epsilon)$ for 2D and 3D.}%
\label{fig: 2d_3d lam_eps}%
\end{figure}

\section{Numerical computations}

\begin{figure}[ptb]
\includegraphics[height=5.5cm]{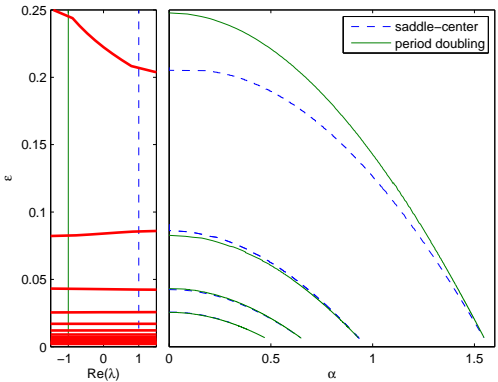}
%\caption[]{2D: Wedges of stability in the parameter space}
\caption{(Color online) 2D. Left: real part of eigenvalue $\lambda$ (bold) at
$\alpha=0$. Right: Wedges of stability in the parameter space. }%
\label{fig: 2d parameter}%
\end{figure}

\begin{figure}[ptb]
\includegraphics[height=5.5cm]{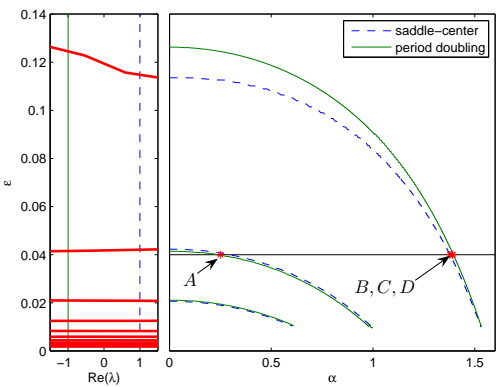}
%\caption[]{2D: Wedges of stability in the parameter space}
\caption{(Color online) 3D. Left: real part of eigenvalue $\lambda$ (bold) at
$\alpha=0$. Right: Wedges of stability in the parameter space. See
FIG.\ref{fig: 3d phase} for phase portraits the parameter values corresponding
to A-D. }%
\label{fig: 3d parameter}%
\end{figure}From the analysis of \cite{turk03} we expect that the stability of
$\gamma(t,\epsilon,\alpha)$ will depend non-trivially on both $\epsilon$ and
the geometrical parameter of the billiard $\alpha$ and that near $\alpha
_{k}=\frac{\pi}{k}$ islands will appear (the limit $\alpha\rightarrow0$ at
which the billiard is not a Sinai billiard, and thus billiard orbits may be
trapped for arbitrarily large number of reflections near the corner has not
been studied in \cite{turk03}). We find that all the regions in the $\left(
\alpha,\epsilon\right)  $ plane at which islands of stability associated with
$\gamma(t,\epsilon,\alpha)$ exist (other islands of stability may co-exist),
emerge from $\alpha=0$ at some finite $\epsilon_{k}^{\pm}$ values, and
converge towards $\left(  \alpha,\epsilon\right)  \rightarrow(\frac{\pi}%
{k},0)$. Hence, we first find the stability of $\gamma(t,\epsilon,\alpha=0)$
by computing the eigenvalues of the monodromy matrix of the return map to the
local cross-section at $x=0$ for a range of $\epsilon$ values. Since there is
always a pair of neutral eigenvalues corresponding to the flow direction, for
the 2d case the monodromy matrix has the eigenvalues $\{1,1,\lambda,\frac
{1}{\lambda}\}$ where $\lambda$ is the largest eigenvalue which is different
from 1. In the 3d case, due to the symmetric form of the geometry, the
spectrum is of the form $\{1,1,\lambda,\frac{1}{\lambda},\lambda,\frac
{1}{\lambda}\}.$(i.e. saddle-foci do not appear). In FIG.
\ref{fig: 2d_3d lam_eps} the real part of $\lambda$ is shown for a range of
$\epsilon$ values for the 2d and 3d cases. The large oscillations from
positive to negative values guarantee the existence of intervals of $\epsilon$
at which \ $\operatorname{Re}\{\lambda\}\in(-1,1)$ - on these intervals
$\lambda$ is imaginary and belongs to the unit circle. In the left panels of
FIG.\ref{fig: 2d parameter} and FIG.\ref{fig: 3d parameter} we present an
enlarged segment of FIG. \ref{fig: 2d_3d lam_eps} with a regular $\epsilon$
scale. These calculations are used to find the values of $\epsilon
=\epsilon_{k}^{\pm}$ at which $\operatorname{Re}\{\lambda\}=\pm1$, where a
saddle-center and a period doubling bifurcations occurs respectively (in the
three dimensional case these are double-bifurcation points due to the
symmetry). Then, starting at $\left(  \alpha,\epsilon\right)  =\left(
0,\epsilon_{k}^{\pm}\right)  $, we use a continuation method for finding the
bifurcation curves for $\alpha>0$, as shown in the right panels of
FIG.\ref{fig: 2d parameter} and FIG.\ref{fig: 3d parameter}. In the wedges
enclosed by these two curves the periodic orbit $\gamma(t,\epsilon,\alpha)$ is
elliptic, with Flouqet multipliers $\exp(\pm i\omega)$ (in the three
dimensional case each multiplier has multiplicity two), and $\omega$ varies
between $0$ and $\pi$ as the wedges are crossed. One expects that this linear
stability will also result in nonlinear stability for most (non-resonant)
$\omega$ values. More elaborate study of the resonances and the relation to
the analytic predictions of \cite{turk03} are of interest but are beyond the
scope of the current paper. For the two dimensional case, we verified that
indeed the phase portraits one obtains as a wedge of stability is crossed are
the familiar islands which appear near a saddle-center and a Hamiltonian
period-doubling bifurcations (e.g. as in the Hamiltonian H\'{e}non map).

%\begin{figure}[ptb]
%\includegraphics[height=5.5cm]{2D_bifur.eps} %\caption[]{2D: Wedges of stability in the parameter space}
%\caption{2D. Left: real part of eigenvalue $\lambda_{3}$ (bold) at $\alpha=0$.
%Right: Wedges of stability in the parameter space. }%
%\label{fig: 2d parameter}%
%\end{figure}
%\begin{figure}[ptb]
%\includegraphics[height=5.5cm]{3D_bifur.eps} %\caption[]{2D: Wedges of stability in the parameter space}
%\caption{3D. Left: real part of eigenvalue $\lambda_{3}$ (bold) at $\alpha=0$.
%Right: Wedges of stability in the parameter space. See FIG.\ref{fig: 3d phase}
%for phase portraits the parameter values corresponding to A-D. }%
%\label{fig: 3d parameter}%
%\end{figure}

\begin{figure*}[ptb]
%\begin{minipage}[t]{5.5cm}
%\includegraphics[height=5.5cm]{ps_y_al_025_ep_004.eps}
%\end{minipage}\\
%\begin{minipage}[t]{5.5cm}
%\includegraphics[height=5.5cm]{ps_y_al_1378_ep_004.eps}
%\end{minipage}
%\begin{minipage}[t]{5.5cm}
%\includegraphics[height=5.5cm]{ps_y_al_139_ep_004.eps} %\caption[]{2D: Wedges of stability in the parameter space}
%\end{minipage}
%\begin{minipage}[t]{5.5cm}
%\includegraphics[height=5.5cm]{ps_y_al_1392_ep_004.eps} %\caption[]{2D: Wedges of stability in the parameter space}
%\end{minipage}
\includegraphics[height=10cm]{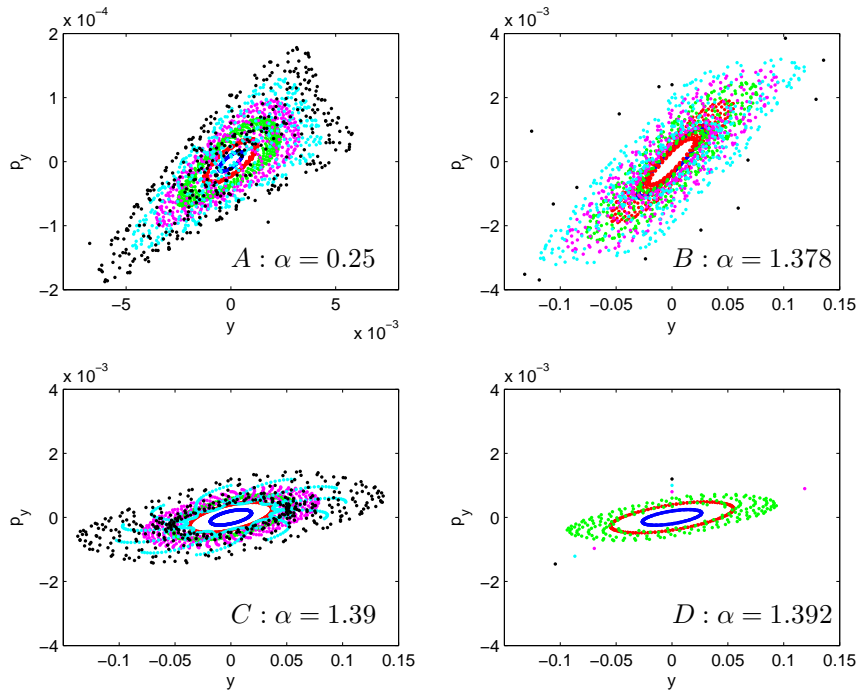}
\caption{(Color online) 3D. Phase portraits $(y,p_{y})$ at cross-section
$x=0,p_{x}>0$ for different values of $\alpha$,$\epsilon=0.04$, see also
FIG.\ref{fig: 3d parameter}. Notice the different scales for the first
stability wedge (B-D) and the second stability wedge (A).}%
\label{fig: 3d phase}%
\end{figure*}

In the three dimensional case, for all $\omega$ values, the multipliers are in
$1:1$ resonance due to the symmetry. For generic systems, for almost all
$\omega$ values (values which are non-resonant with the frequency of
$\gamma(t,\epsilon,\alpha)$), we expect to have non-linear stability (see e.g.
\cite{MSB86}). Indeed, projections of the four dimensional symplectic return
map to $x=0$ for several $(\alpha,\epsilon)$ values are shown in
FIG.\ref{fig: 3d phase}. It is demonstrated that indeed inside the wedged
region $\gamma(t,\epsilon,\alpha)$ is nonlinearly stable for the full
integration time (approximately 4000 periods). Moreover, if we add a
sufficiently small, a-symmetric perturbation to the potential (e.g.
$V=W+\delta\cos(y+\eta)\cos(z+\mu)$ with $\delta,\eta,\mu=O(0.0001)$) we find
that the effective stability region still persists. For the phase-space
simulations we use a symplectic integrator (GniCodes \cite{HH03}), which keeps
$h$ up to accuracy of $10^{-11}$. Thus, we can confidently detect islands with
transversal kinetic energy of up to $10^{-8}$ (so $(p_{y},p_{z})=O(10^{-4})$).
This limits our phase-space calculations to $\varepsilon\approx0.04$ --
smaller values of $\varepsilon$ produce smaller islands and their detection
via phase space plots requires a higher accuracy in the integration. We stress
though that the calculations of the bifurcation curves are accurate for much
smaller $\varepsilon$ values; in these calculation only a single return map is
computed and there exists a sharp transition between large positive and large
negative values of the eigenvalues (see left panels of FIG.
\ref{fig: 2d parameter},\ref{fig: 3d parameter}), so the existence of elliptic
regimes is guaranteed. Comparing the 2d and 3d wedges of stability it appears
that the 3d wedges are indeed narrower.

\section{Concluding remarks}

While the appearance of islands in two-degrees of freedom steep Hamiltonian
systems is somewhat expected, the mechanisms for their appearance in the
higher dimensional settings is not as well understood (see \cite{MSB86,GST04}
for some generic possibilities). Furthermore, their appearance guarantees only
effective stability due to the possible existence of Arnold diffusion
\cite{GDFG89}. Nonetheless, by KAM theory, in the non-degenerate case, a large
set of initial conditions belongs to KAM tori and thus stay forever near the
stable periodic orbit. Thus, the existence of islands in the higher
dimensional setting \emph{implies that ergodicity is destroyed} independently
of the possible leakage out of the effective stability zone after an
exponentially long time. This latter possibility suggests that stickiness may
be an interesting event also in this higher dimensional setting.

Here, we propose for the first time a mechanism for the creation of stability
islands for smooth systems which are arbitrarily close to strictly dispersing
three dimensional billiards; we showed that potentials $V(q;\epsilon,\alpha)$
that become arbitrarily steep as $\epsilon\rightarrow0,$ possess wedges in the
$(\epsilon,\alpha)$-plane at which a periodic orbit is elliptic. Thus, on one
hand, there exist one-parameter families of potentials $V(q;\epsilon
,\alpha(\epsilon))$ which have a stable periodic orbit for arbitrarily small
$\epsilon$. Since we showed that in the wedges $\alpha(\epsilon)\rightarrow
\alpha(0)>0$ as $\epsilon\rightarrow0$, it follows that these potentials have
islands of stability even when they are \emph{arbitrarily close to a hard wall
dispersing (Sinai) billiards. }On the other hand, for any fixed $\alpha
\in(0,\frac{\pi}{2})$ there exists an interval of positive $\epsilon$ values
for which islands of stability exist. Thus, these islands may be destroyed by
either making the potential steeper OR softer -- a somewhat non-intuitive result.

\section{Acknowledgment}

We thank U. Smilansky and D. Turaev for discussions and comments. We
acknowledge the support of the Israel Science Foundation (Grant 926/04) and
the Minerva foundation.

\bibliographystyle{amsplain}
\providecommand{\bysame}{\leavevmode\hbox
to3em{\hrulefill}\thinspace}
\providecommand{\MR}{\relax\ifhmode\unskip\space\fi MR }
\providecommand{\MRhref}[2]{%
  \href{http://www.ams.org/mathscinet-getitem?mr=#1}{#2}
} \providecommand{\href}[2]{#2}

\end{document}